\documentclass[12pt,a4paper]{article}
\usepackage{amssymb}
\usepackage{amsmath}

\begin{document}

\thispagestyle{plain}

%\rightline{Physics/9912008}

\author{N. T. Anh\thanks{%
Present address: {\it Institute for Nuclear Science and Technique, Hanoi,
Vietnam.}} \thanks{%
Email: anhnt@vol.vnn.vn} \\
%EndAName
\\
{\it Department of Theoretical Physics, Hanoi National University,}\\
{\it \ High College of Physics, Institute of Theoretical Physics,}\\
{\it \ Hanoi , Vietnam.}}
\title{CAUSALITY \\
The Nature of Everything \\
\bigskip \bigskip \medskip }
\date{(1991)}
\maketitle

\begin{abstract}
\parskip5pt

We pursue research leading towards the nature of causality in the universe.
We establish the equation of the universe's evolution from the
universe-state function and its series expansion, in which causes and
effects connect together to construct a linked chain of causality. And the
equation of causality [1] is rederived. Therefrom, our theory informs the
progress of the universe and, simulaneously, a law that presents everywhere.
Furthermore, we lay some foundations for a new mathematical phrasing of the
movement of physics. That is the life-like mathematics - a mathematics full
of life.
\end{abstract}

\newpage

\pagestyle{myheadings} \markright{CAUSALITY: The Nature of Everything -- N. T. Anh}

\parskip 9pt

\baselineskip18pt

\section{INTRODUCTION}

To pursue the idea of the previous article [1] a question arises that if the
law of causality is the most general law, then about what will it inform us
in the whole of the universe, in the birth, expansion and conclusion of the
universe, as well as in some concrete phenomenon and process in the universe?

The transformation chain ``{\it Difference} $\rightarrow $ {\it Contradiction%
} $\rightarrow $ {\it Solving of contradiction}'' is also the expansion
chain of the universe. The expansion chain of the universe starts from the
absolute infinite and homogeneous space. In the absolute space there is a
Nothing, but itself is the one, is the unique: the content is zero, the form
is one. Thus as acting negatively all the absolute space acts negatively
just its unique existence. Therefrom the immanent contradiction of this
state of the universe is infinitely great, and the universe self-solves by
expanding into infinite series of smaller contradictions and solving them.
The consequence is that our world was born.

In order to describe quantitatively all processes of the birth, expansion
and conclusion of the universe, we establish a universal equation from the
universe-state function in Sec. 2. And the equation of causality is anew
devised exactly in Sec. 3. Our conclusion is given in Sec. 4. In Appendices
A and B we present a few foundations for functor and life-like mathematics.

\section{UNIVERSAL EQUATION}

{\bf 1!}\quad At the debut of each expansion of the universe, the $%
f\hspace*{0cm}\hspace*{-0.1cm}!$ function characterizing for the early state
of the universe is equal gradually to zero everywhere and forming the
absolute vacuum-state function $f\hspace*{-0.1cm}!(~)$. 
\[
f\hspace*{-0.1cm}!=f\hspace*{-0.1cm}!(~). 
\]

Continuously, in this expansion, the vacuum state of the universe becomes
the absolute space state but still is always equal to zero everywhere,
because in the vacuum of the universe there contain infinitely many absolute
spaces 
\[
f\hspace*{-0.1cm}!(~)=f\hspace*{-0.1cm}!(K), 
\]
where $f\hspace*{-0.1cm}!(K)$ is called absolute space-state function of the
set of absolute spaces in the universe's vacuum, with $K$ being absolute
space set $K=\{...k...\}$.

Therefore, the absolute space-state function satisfies the following
condition 
\begin{equation}
\left\{ 
\begin{array}{l}
f\hspace*{-0.1cm}!(K)=0,\qquad \forall k\in K \\ 
\int f\hspace*{-0.1cm}!(K)dk=1,
\end{array}
\right.  \tag{1.1}
\end{equation}
where the notation ``0'' is the ``nothing'' in the universe's vacuum, and
``1'' symbolizes the unique existence of the universe's vacuum.

$f\hspace*{-0.1cm}!(K)$ is equal to zero everywhere, but its integral with
respect to space is unique, or the whole of $f\hspace*{-0.1cm}!(K)$ over
space is unique. (Right in whole integral region, $f\hspace*{-0.1cm}!(K)$
still is equal to zero - The absolute vacuum has nothing, but has uniquely
that ``nothing''.)

$f\hspace*{-0.1cm}!(K)$ is limitless and homogeneous everywhere, is
determined in every neighborhood of point spaces $\kappa _{i}$ in a set of
absolute point spaces ${\cal K}(\kappa _{i}\in {\cal K}=\{...\kappa ...\})$,
is continuous with $k_{i}=\kappa _{i}$, and exists the limit $%
\lim_{k_{i}\rightarrow \kappa _{i}}$ $f\hspace*{-0.1cm}!(K)=$ $%
f\hspace*{-0.1cm}!({\cal K})$. $f\hspace*{-0.1cm}!(K)$ is continuously
differentiable on the set of point spaces ${\cal K}$, in which all direct
and reverse partial derivatives to all orders exist and are continuous due
to the existence of limits, respectively.

\bigskip

{\bf 2!}\quad The state functions of the absolute vacuum as well as of the
absolute space determine the states of the homogeneous state vacuum and the
homogeneous space, in which as transferring with respect to all degrees of
freedom there cannot be any arbitrary immanent difference.

But, the absolute vacuum is a unique and homogeneous state in which there is
nothing. As self-acting negatively, it leads to self-acting positively
itself and vice versa. In the unique state there contains nothing but still
exists a ``nothingness''! For that reason, in overall the absolute vacuum
still contains a difference: between the ``nihility'' and the unique state
of this ``nihility''. That difference is infinitely great, and therefore the
immanent contradiction of this state is also infinitely great. It is easy to
be identified this contradiction in the relation (1.1).

So the self-solving is performed by expanding $f\hspace*{-0.1cm}!(K)$ into
series in the whole of space, 
\begin{eqnarray*}
\left( 
\begin{array}{cccc}
& \vdots & \vdots &  \\ 
\cdots & A^{(0)}(k) & \sum\limits_{h}A_{h}^{(+1)}(k) & \cdots \\ 
\cdots & \sum\limits_{h}A_{h}^{(-1)}(k) & A^{(0)}(k) & \cdots \\ 
& \vdots & \vdots & 
\end{array}
\right) &=&\left( 
\begin{array}{cccc}
& \vdots & \vdots &  \\ 
\cdots & A^{(0)}(k) & \sum\limits_{h}\left. A_{h}^{(+1)}(k)\right| _{\kappa
} & \cdots \\ 
\cdots & \sum\limits_{h}\left. A_{h}^{(-1)}(k)\right| _{\kappa } & 
A^{(0)}(k) & \cdots \\ 
& \vdots & \vdots & 
\end{array}
\right) \\
&&\times \left( 
\begin{array}{cccc}
& \vdots & \vdots &  \\ 
\cdots & 1 & \sum\limits_{h}\int_{\kappa _{h}^{-}}^{k_{h}^{-}}d\varkappa
_{h}^{-} & \cdots \\ 
\cdots & \sum\limits_{h}\int_{\kappa _{h}^{+}}^{k_{h}^{+}}d\varkappa
_{h}^{+} & 1 & \cdots \\ 
& \vdots & \vdots & 
\end{array}
\right) ,
\end{eqnarray*}
or in the general form, 
\begin{equation}
f\hspace*{-0.1cm}!(K)=\left. f\hspace*{-0.1cm}!(K)\right| _{{\cal K}}S_{%
{\cal K}}^{K}  \tag{1.2}
\end{equation}
under the space reflection, where $f\hspace*{-0.1cm}!(K)$, $\left.
f\hspace*{-0.1cm}!(K)\right| _{{\cal K}}$, and $S_{{\cal K}}^{K}$ contain
infinitely many variable elements.

Conversely, if there is a linear transfer variation from $K$ space to ${\cal %
K}$ space, then 
\[
\left. f\hspace*{-0.1cm}!(K)\right| _{{\cal K}}=f\hspace*{-0.1cm}!(K)S_{K}^{%
{\cal K}}, 
\]
where $S_{K}^{{\cal K}}$ is different from $S_{{\cal K}}^{K}$ by permuting
of superior and inferior limits of integrals.

For that reason, if doing an action $\left. f\hspace*{-0.1cm}!(K)\right| _{%
{\cal K}}$ onto $S_{{\cal K}}^{K}$, then 
\begin{equation}
S_{K}^{{\cal K}}S_{{\cal K}}^{K}=S_{{\cal K}}^{K}S_{K}^{{\cal K}}=I 
\tag{1.3}
\end{equation}
with $I$ unit.

\bigskip

{\bf 3!}\quad Now, we return to consider Eq. (1.2). If instead of $K$ we let 
$(K-{\cal K})$, then $f\hspace*{-0.1cm}!(K)$ becomes 
\begin{equation}
f\hspace*{-0.1cm}!(K-{\cal K})=\left. f\hspace*{-0.1cm}!(K-{\cal K})\right|
_{(K-{\cal K})=0}S_{(K-{\cal K})=0}^{(K-{\cal K})},  \tag{1.4}
\end{equation}

And if we define $\int_{\kappa _{i}}^{k_{i}}d\varkappa _{i}=M_{i}$, then $S_{%
{\cal K}}^{K}={\cal M}$, namely 
\[
\left( 
\begin{array}{cccc}
& \vdots & \vdots &  \\ 
\cdots & 1 & \sum\limits_{h}\int_{\kappa _{h}^{-}}^{k_{h}^{-}}d\varkappa
_{h}^{-} & \cdots \\ 
\cdots & \sum\limits_{h}\int_{\kappa _{h}^{+}}^{k_{h}^{+}}d\varkappa
_{h}^{+} & 1 & \cdots \\ 
& \vdots & \vdots & 
\end{array}
\right) =\left( 
\begin{array}{cccc}
& \vdots & \vdots &  \\ 
\cdots & 1 & \sum\limits_{h}\frac{M_{h}^{-}}{-1!} & \cdots \\ 
\cdots & \sum\limits_{h}\frac{M_{h}^{+}}{+1!} & 1 & \cdots \\ 
& \vdots & \vdots & 
\end{array}
\right) . 
\]

Thus, Eq. (1.4) will be 
\begin{equation}
f\hspace*{-0.1cm}!({\cal M})=\left. f\hspace*{-0.1cm}!({\cal M})\right| _{%
{\cal M}=0}{\cal M}  \tag{1.5}
\end{equation}
where each element equation has the form as a functor 
\[
A(M)=e^{\Sigma M\widehat{A}}\left. A(M)\right| _{M=0}. 
\]

From calculating $S_{{\cal K}}^{K}$, we are convenient to consider the
condition (1.3). After we define $\int_{k_{i}}^{\kappa _{i}}d\varkappa
_{i}=-\int_{\kappa _{i}}^{k_{i}}d\varkappa _{i}=-M_{i}$, then $S_{K}^{{\cal K%
}}={\cal M}^{-1}$, namely 
\[
\left( 
\begin{array}{cccc}
& \vdots & \vdots &  \\ 
\cdots & 1 & \sum\limits_{h}\int_{k_{h}^{-}}^{\kappa _{h}^{-}}d\varkappa
_{h}^{-} & \cdots \\ 
\cdots & \sum\limits_{h}\int_{k_{h}^{+}}^{\kappa _{h}^{+}}d\varkappa
_{h}^{+} & 1 & \cdots \\ 
& \vdots & \vdots & 
\end{array}
\right) =\left( 
\begin{array}{cccc}
& \vdots & \vdots &  \\ 
\cdots & 1 & \sum\limits_{h}\frac{-M_{h}^{-}}{-1!} & \cdots \\ 
\cdots & \sum\limits_{h}\frac{-M_{h}^{+}}{+1!} & 1 & \cdots \\ 
& \vdots & \vdots & 
\end{array}
\right) . 
\]

Thus, in order for ${\cal MM}^{-1}={\cal M}^{-1}{\cal M}=I$ then 
\begin{equation}
M^{+}M^{-}=M^{-}M^{+}=0,  \tag{1.6}
\end{equation}
where $M^{+}$ and $M^{-}$ are some contradictions which belong to the two
reflective worlds, and notice here the formula 
\[
\sum\limits_{i=0}^{n}\frac{(-)^{i}}{(n-i)!i!}=\frac{1}{n!}%
\sum\limits_{i=0}^{n}(-)^{i}\left( 
\begin{array}{c}
n \\ 
i
\end{array}
\right) =0,
\]
with meaning that the sum of generation and annihilation quantities is
constant (conservation of quanta).

\bigskip

{\bf 4!}\quad From the second in the condition (1.1) of the absolute
space-state function, we consider in the set of point-spaces ${\cal K}$, 
\[
1=\int_{\kappa }\hspace{-0.37cm}\bullet \ f\hspace*{-0.1cm}!(K)d\varkappa
=f\hspace*{-0.1cm}!(K)\left. A^{(0)}(k)\right| _{\kappa }, 
\]
or can write 
\[
1=f\hspace*{-0.1cm}!(K-{\cal K})\left. A^{(0)}(k-\kappa )\right| _{(k-\kappa
)=0},\text{ i.e.} 
\]
\begin{equation}
1=f\hspace*{-0.1cm}!({\cal M})\left. A^{(0)}(M)\right| _{M=0}.  \tag{1.7}
\end{equation}
And consider in an interval of integration from the set of point-spaces $%
{\cal K}$ to the set of absolute spaces $K$%
\[
1=\int_{\kappa }^{k}f\hspace*{-0.1cm}!(K)d\varkappa =f\hspace*{-0.1cm}!(K-%
{\cal K})[k-\kappa ]^{(0)},\text{ i.e.} 
\]
\begin{equation}
1=f\hspace*{-0.1cm}!({\cal M})M^{(0)}.  \tag{1.8}
\end{equation}

Following the life-like mathematics, let (1.7) and (1.8) sink into together
we obtain an equivalent relation 
\begin{equation}
\left. A^{(0)}(M)\right| _{M=0}=M^{(0)}.  \tag{1.9}
\end{equation}

After we differentiate partially (in direct and reverse orientations) to all
orders with respect to components in the two elements $\left.
A^{(0)}(M)\right| _{M=0}$ and $M^{(0)}$, respectively, then (1.9) may be
written as the form 
\[
\left( 
\begin{array}{cccc}
& \vdots & \vdots &  \\ 
\cdots & \left. A^{(0)}(M)\right| _{M=0} & \sum\limits_{h}\left.
A_{h}^{(+1)}(M)\right| _{M=0} & \cdots \\ 
\cdots & \sum\limits_{h}\left. A_{h}^{(-1)}(M)\right| _{M=0} & \left.
A^{(0)}(M)\right| _{M=0} & \cdots \\ 
& \vdots & \vdots & 
\end{array}
\right) =\left( 
\begin{array}{cccc}
& \vdots & \vdots &  \\ 
\cdots & M^{(0)} & \sum\limits_{h}a_{h}^{+}M_{h}^{(+1)} & \cdots \\ 
\cdots & \sum\limits_{h}a_{h}^{-}M_{h}^{(-1)} & M^{(0)} & \cdots \\ 
& \vdots & \vdots & 
\end{array}
\right) . 
\]

Namely, 
\begin{equation}
\left. f\hspace*{-0.1cm}!({\cal M})\right| _{{\cal M}=0}={\cal M}^{\prime },
\tag{1.10}
\end{equation}
where the factors $a$ are degree-of-freedom transfer coefficients.

Therefrom, we have the following equation, from Eq. (1.5), 
\begin{equation}
f\hspace*{-0.1cm}!({\cal M})={\cal M}^{\prime }{\cal M}.  \tag{1.11}
\end{equation}
This equation is named universal equation, where each element is of the form 
\[
A(M)=e^{\Sigma aM\widehat{M}}M. 
\]

The two spaces $k_{i}$ and $\kappa _{i}$ in the respective sets $K$ and $%
{\cal K}$ are different from each other, but the $\kappa _{i}$ space lies in
the $k_{i}$ space. Id est, these two spaces differ mutually but they concern
directly with each other through limiting of integral sum, so they
annihilate mutually generating the contradiction between them 
\[
\lbrack k_{i}-\kappa _{i}]=M_{i}, 
\]
where ``$-$'' is signed annihilation operation between two spaces $k_{i}$
and $\kappa _{i}$. The dimension of contradiction depends on the dimension
of spaces $k_{i}$ and $\kappa _{i}$ which contribute to create that
contradiction.

Thus, (1.9) means that $\left. A^{(0)}(k)\right| _{\kappa }$ at the point
space $\kappa $ of the set ${\cal K}$ is just the contradiction $[k-\kappa
]=M$ between the absolute space $k$ of the set $K$ and the point space $%
\kappa $ of the set ${\cal K}$ contained in $K$. For the reason that $\left.
A^{(0)}(k)\right| _{\kappa }$ reflects right the contradiction between the $%
\kappa $ space for which this function characterizes and the $k$ space,
characterized by $A(k)$ that is rather conservative than $\left. A(k)\right|
_{\kappa }$ (in immanent contradiction).

Hence, partial derivatives of $\left. A(k)\right| _{\kappa }$ to all orders
with respect to all possible quantities in point spaces $\kappa _{i}$ of the
set ${\cal K}$ are just partial derivatives of the contradiction $M$ to all
orders with respect to all possible degrees of freedom in spaces (i.e. with
respect to degrees of transfer of contradiction), with $a$ being the
degree-of-freedom transfer coefficients. And then, the derivatives ${\cal M}%
^{\prime }$ of contradiction with respect to contradiction-transfer
facilities become the expedients for solving of contradiction $\widehat{%
{\cal M}}$.

In ${\cal M}$, the unit elements in the main diagonal form a reflecting
mirror demarcating the two worlds. When an element is generated in one
world, a some element in the other is simultaneously annihilated, and
conversely. (they are conserved of degree). The two reflecting worlds seem
to be alike but if an arbitrary element of the one unites some arbitrary
element of the other, they both will annihilate mutually to zero, Eq. (1.6).
The annihilation to zero from two some elements of the two reflecting worlds
is to satisfy the condition (1.3).

Each element of one world will become a some element of the other only after
it has passed the reflecting mirror, the reflecting mirror has annihilated
all natures of old world in this element so that it can receive new natures
when it enter into a new world.

\bigskip

{\bf 5!}\quad Let us now generalize all above concepts. If the universe
comprises a set of actions, thereat we can regard that $K$ is set of actions
which exist in all absolute spaces and ${\cal K}$ is the set of actions
which exist in all point spaces. $f\hspace*{-0.1cm}!(~)$ is the state
function of the zero action, $f\hspace*{-0.1cm}!(K)$ is the action function
of absolute spaces, and $\left. f\hspace*{-0.1cm}!(K)\right| _{{\cal K}}$ is
the action function of point spaces. They satisfy the condition (1.1), with
``0'' being ``zero action'', and ``1'' being the action of its unique
existence.

A some contradiction, thereat, will be defined as the simultaneous
coexistence of two mutual annihilation actions, 
\[
M_{i}=[k_{i}-\kappa _{i}],\qquad k_{i}\in K\text{ and }\kappa _{i}\in {\cal K%
}, 
\]
and satisfies Eq. (1.6).

From the condition of the uniqueness of the zero action, we obtain the
universal equation (1.11) where ${\cal M}^{\prime }$ contains elements as
derivatives of contradictions with respect to all possible degrees of
freedom. On the other hand, they act onto contradictions making
contradictions to vary with respect to some degrees of freedom -- i.e.
making contradictions to be solved over some facilities for transfer of
contradiction, so derivatives of contradictions with respect to all degrees
of freedom are just contradiction-solving expedients over all
contradiction-transfer facilities.

The universal equation manifests the birth, evolution and conclusion of the
universe, beginning from the absolute vacuum and then its expanding out the
whole of the universe to infinitely many repeat period chains. The universal
equation expresses just solving an infinitely great immanent contradiction
of the absolute vacuum state by dividing into infinitely many small
contradictions to solve them.

In the absolute vacuum there are infinitely many point spaces, solving
contradictions between the vacuum and point spaces makes point spaces
concentrate and crystallized together (accretion) forming early matters.
They combine together, and continuous so, gradually forming the world in a
unity entity of the nature.

If speaking in the geometrical language, contradictions between the zero
curvature of the absolute vacuum (or of a sphere with an infinitely large
radius) and infinitely high curvatures of point spaces (or of spheres with
infinitely small radii) must vary to reach states with lowest immanent
contradictions, i.e. curvatures must vary to reach and conclude some
curvatures (following tendency). The curvature of the absolute vacuum
continuously conserves and is equal to zero, so these contradictions always
are solved with respect to the tendency\ of decreasing curvatures of point
spaces to zero (i.e. increasing their radii to infinity) so as to coincide
with the curvature of the absolute vacuum. As curvatures of point spaces
decrease just as point spaces concentrate and crystallize together to form
new spaces with lower curvatures (i.e. with larger radii).

Concreting that at each point in the absolute vacuum there is an appearance
of infinitely many centripetal flows (inflow). Isophasic flows force to
appear an impulse at this point, immediately, that impulse causes an
immanent contradiction and the contradiction is solved directly by emitting
backward flows (outflow) to decrease strength of inward flows. Then after
that, emitter is weak gradually and a new cycle will begin, etc.

In the absolute vacuum such infinitely many point spaces create a global
motion of the universe generating infinitely many thermal currents
everywhere in the universe. The universe becomes full of life. Motion of
infinitely many point spaces leads to the formation of groups separated by
isophasic current points. (Maybe the opposite phase groups would create the
positive and negative extrema or mutually symmetrical extrema. And points
which have intermediary \ phases would disperse to form different groups
then move gradually to one of extrema). Each group creates one proper wave
of flows giving rise to the universe to ``breathe'' lively.

In the universal equation, the sign ``$=$'' has the meaning that: at the
debut of each expansion process, the equation performs generally in the
forward direction (rightwards), after that the equation is in relative
equilibrium when there are new contradictions generated and there also are
contradictions solved, finally the equation performs generally in the
backward direction (leftwards) when contradictions are solved more and more
-- at that time the expansion series converges uniformly to $%
f\hspace*{-0.1cm}!(K)=f\hspace*{-0.1cm}!(~)=f\hspace*{-0.1cm}!$.

Each numerical value under contradictions is determined by reflecting them
onto the quantity $0!$ of the absolute vacuum, and as contradictions vary
the numerical values also alter after. The state function thus expands into
chains in the reversible process, contradictions are generated gradually,
and quantities also gradually appear more and more. In element equations,
when expanding sums there will be chains of terms, each chain is a period,
later chain is in higher development degree than sooner chain.

In a few real concrete cases, the expansion of the equation requires
strictly to determine superior limits of sums -- i.e. to obey the rule of
filling ``reservoirs'' of the universe. If the equation expands from small
contradictions to great contradictions, then when a some contradiction which
corresponds with a superior limit index of a sum is greater than other
contradiction which corresponds with an inferior limit index of a successive
sum, the smaller contradiction will be expanded foremost. This is similar to
the expansion with respect to levels of energy minimization of elements in
the periodic system.

Thus, all phenomena in the universe seem to proceed in an accomplished order
and the whole universe continuously obeys a law. A some term of series does
not really open out yet but it has denoted a progress orientation of future
expansion.

About application, above all the universal equation is used for deriving
directly the equation of causality, after that for building equations of
many quantity sets in each scope of a researched space and for researching
constructions and systems of worlds in the universe.

\section{EQUATION OF CAUSALITY}

We now research the reciprocal reflection from the space world or the world
of actions to the time world. The time world stands of a fundamental
background or field, onto which other worlds all reflect as a whole.
Therefore the time world is relatively independent, and other worlds seem to
be timeworld-dependent. In the time world the process of variation and
transformation also is performed and formed from the functions $%
f\hspace*{-0.1cm}!$ and $f\hspace*{-0.1cm}!(~)$ and expanded into series,
finally reaches the homogeneous state and closes a reflection period.

As the world of actions reflects on the time world, the set of actions $K$
is dependent on the set of times $T=\{...\top ...\}$ which exist in absolute
spaces, and ${\cal K}$ is dependent on ${\cal T}=\{...\tau ...\}$ in point
spaces. Thereby, $f\hspace*{-0.1cm}!(~)$ reflects on time forming the
zero-action function of the absolute vacuum in the non-time state, $%
f\hspace*{-0.1cm}!(K)$ becomes the function $f\hspace*{-0.1cm}!(K_{(T)})$;
and $\left. f\hspace*{-0.1cm}!(K)\right| _{{\cal K}}$ grows $\left.
f\hspace*{-0.1cm}!(K_{(T)})\right| _{{\cal K}_{({\cal T})}}$.

Similarly, in the time world the function $f\hspace*{-0.1cm}!(K_{(T)})$ also
is expanded into series by linear transformation from the time ${\cal T}$ to 
$T$, 
\begin{equation}
f\hspace*{-0.1cm}!(K_{(T)})=\left. f\hspace*{-0.1cm}!(K_{(T)})\right| _{%
{\cal K}_{({\cal T})}}S_{{\cal T}}^{T},  \tag{2.1}
\end{equation}
or may be written (in the time inflection) 
\begin{eqnarray*}
\left( 
\begin{array}{cccc}
& \vdots  & \vdots  &  \\ 
\text{{\small ..}} & A^{(0)}(k_{(\top )}) & \sum\limits_{h}A_{\top
_{h}}^{(+1)}(k_{(\top )}) & \text{{\small ..}} \\ 
\text{{\small ..}} & \sum\limits_{h}A_{\top _{h}}^{(-1)}(k_{(\top )}) & 
A^{(0)}(k_{(\top )}) & \text{{\small ..}} \\ 
& \vdots  & \vdots  & 
\end{array}
\right) \hspace{-0.05cm}\hspace{-0.1cm} &=&\hspace{-0.05cm}\hspace{-0.1cm}%
\left( 
\begin{array}{cccc}
& \vdots  & \vdots  &  \\ 
\text{{\small ..}} & A^{(0)}(k_{(\top )}) & \sum\limits_{h}\left. A_{\top
_{h}}^{(+1)}(k_{(\top )})\right| _{\kappa _{(\tau )}} & \text{{\small ..}}
\\ 
\text{{\small ..}} & \sum\limits_{h}\left. A_{\top _{h}}^{(-1)}(k_{(\top
)})\right| _{\kappa _{(\tau )}} & A^{(0)}(k_{(\top )}) & \text{{\small ..}}
\\ 
& \vdots  & \vdots  & 
\end{array}
\right)  \\
&&\times \left( 
\begin{array}{cccc}
& \vdots  & \vdots  &  \\ 
\text{{\small ...}} & 1 & \sum\limits_{h}\int_{\tau _{h}^{-}}^{\top
_{h}^{-}}dt_{h}^{-} & \text{{\small ...}} \\ 
\text{{\small ...}} & \sum\limits_{h}\int_{\tau _{h}^{+}}^{\top
_{h}^{+}}dt_{h}^{+} & 1 & \text{{\small ...}} \\ 
& \vdots  & \vdots  & 
\end{array}
\right) .
\end{eqnarray*}

If instead of $K_{(T)}$ we write $[K_{(T)}-{\cal K}_{({\cal T})}]$, and
after putting $\int_{\tau _{i}}^{\top _{i}}dt_{i}=\Delta t_{i}$, 
\[
S_{{\cal T}}^{T}={\it \Upsilon }, 
\]
with $S_{{\cal T}}^{T}S_{T}^{{\cal T}}={\it \Upsilon \Upsilon }^{-1}=I$
where 
\begin{equation}
\Delta t^{+}\Delta t^{-}=\Delta t^{-}\Delta t^{+}=0,  \tag{2.2}
\end{equation}
or 
\[
\left( 
\begin{array}{cccc}
& \vdots & \vdots &  \\ 
\text{{\small ...}} & 1 & \sum\limits_{h}\int_{\tau _{h}^{-}}^{\top
_{h}^{-}}dt_{h}^{-} & \text{{\small ...}} \\ 
\text{{\small ...}} & \sum\limits_{h}\int_{\tau _{h}^{+}}^{\top
_{h}^{+}}dt_{h}^{+} & 1 & \text{{\small ...}} \\ 
& \vdots & \vdots & 
\end{array}
\right) =\left( 
\begin{array}{cccc}
& \vdots & \vdots &  \\ 
\text{{\small ...}} & 1 & \sum\limits_{h}\frac{\Delta t_{h}^{-}}{-1!} & 
\text{{\small ...}} \\ 
\text{{\small ...}} & \sum\limits_{h}\frac{\Delta t_{h}^{+}}{+1!} & 1 & 
\text{{\small ...}} \\ 
& \vdots & \vdots & 
\end{array}
\right) , 
\]
then Eq. (2.1) will be 
\begin{equation}
f\hspace*{-0.1cm}!({\cal M}_{({\it \Upsilon })})=\left. f\hspace*{-0.1cm}!(%
{\cal M}_{({\it \Upsilon })})\right| _{{\cal M}_{({\it \Upsilon })}=0}{\it %
\Upsilon },  \tag{2.3}
\end{equation}
where each element has the functor form, 
\[
A(M_{(t)})=e^{\Sigma \Delta t\widehat{\stackrel{\text{ }\cdot }{A}}}\left.
A(M_{(t)})\right| _{M_{(t)}=0}. 
\]

From the unique condition of the zero action, similar to the previous
section, we obtain 
\begin{equation}
\left. A^{(0)}(M_{(t)})\right| _{M_{(t)}=0}=M_{(t)}^{(0)}  \tag{2.4}
\end{equation}
in the time world.

After direct and reverse partial differentiating to all orders both of the
side hand of this element with respect to time, respectively, we obtain 
\begin{equation}
\left. f\hspace*{-0.1cm}!({\cal M}_{({\it \Upsilon })})\right| _{{\cal M}_{(%
{\it \Upsilon })}=0}=\stackrel{\text{ }\cdot }{{\cal M}}.  \tag{2.5}
\end{equation}
And we can express it under the form 
\[
\left( 
\begin{array}{cccc}
& \vdots & \vdots &  \\ 
\text{{\small ..}} & A^{(0)}(M_{(t)}) & \sum\limits_{h}\left.
A_{t_{h}}^{(+1)}(M_{(t)})\right| _{M_{(t)}=0} & \text{{\small ..}} \\ 
\text{{\small ..}} & \sum\limits_{h}\left. A_{t_{h}}^{(-1)}(M_{(t)})\right|
_{M_{(t)}=0} & A^{(0)}(M_{(t)}) & \text{{\small ..}} \\ 
& \vdots & \vdots & 
\end{array}
\right) \hspace{-0.1cm}=\hspace{-0.1cm}\left( 
\begin{array}{cccc}
& \vdots & \vdots &  \\ 
\text{{\small ..}} & M^{(0)} & \sum\limits_{h}(-)^{+1}M_{t_{h}}^{(+1)} & 
\text{{\small ..}} \\ 
\text{{\small ..}} & \sum\limits_{h}(-)^{-1}M_{t_{h}}^{(-1)} & M^{(0)} & 
\text{{\small ..}} \\ 
& \vdots & \vdots & 
\end{array}
\right) 
\]
where coefficients $(-)$ arise from the reason that contradiction varies
inversely as time.

Thus, we have 
\begin{equation}
f\hspace*{-0.1cm}!({\cal M}_{({\it \Upsilon })})=\stackrel{\text{ }\cdot }{%
{\cal M}}{\it \Upsilon }.  \tag{2.6}
\end{equation}
This equation is named time-world equation, where each element is of the
form 
\[
A(M_{(t)})=e^{\Sigma (-)\Delta t\widehat{\stackrel{\text{ }\cdot }{M}}%
}M_{(t)}, 
\]
with the contradiction between the action $k_{(\top )}$ of the set $K_{(T)}$
and the action $\kappa _{(\tau )}$ of the set ${\cal K}_{({\cal T})}$ being $%
M$ and also dependent on the time $t$: $M_{(t)}$ in the set ${\cal M}_{({\it %
\Upsilon })}$.

Therefore, all-order partial derivatives of the function $\left.
A^{(0)}(k_{(\top )})\right| _{\kappa _{(\tau )}}$ with respect to all times
in point spaces are also just all-order partial derivatives of contradiction
with respect to all times, with coefficients $(-)$ generated by reflecting
from the world of actions to the time world. Therefrom, the derivative $%
\stackrel{\text{ }\cdot }{{\cal M}}$ of contradiction with respect to time
becomes just the contradiction-variation rapidity (violence) $\widehat{%
\stackrel{\text{ }\cdot }{{\cal M}}}$.

Conforming to the life-like mathematics, after sinking the universal
equation into the time-world equation we obtain a general universal
equation, 
\begin{equation}
f\hspace*{-0.1cm}!({\cal M}_{({\it \Upsilon })})={\cal M}^{\prime }{\cal M}=%
\stackrel{\text{ }\cdot }{{\cal M}}{\it \Upsilon }.  \tag{2.7}
\end{equation}

Elements of the universal equation unite respective elements of the
time-world equation, where each functor element has the form 
\[
A(M_{(t)})=e^{\Sigma aM_{(t)}\widehat{M}}M_{(t)}=e^{\Sigma (-)\Delta t%
\widehat{\stackrel{\text{ }\cdot }{M}}}M_{(t)}. 
\]

Consider an arbitrary contradiction in ${\cal M}$, we obtain 
\begin{equation}
aM^{\prime }M=-\stackrel{\text{ }\cdot }{M}\Delta t,  \tag{2.8}
\end{equation}
with $M^{\prime }\in {\cal M}^{\prime }$ and $\stackrel{\text{ }\cdot }{M}%
\in \stackrel{\text{ }\cdot }{{\cal M}}$.

From this equation, we identify that $a$ is not only to be a
degree-of-freedom transfer coefficient but also to be a world transfer
coefficient. In this case it is from the space world to the time world.

Supposing that $(M^{\prime }M)$ reflects from the space world to the time
world, 
\[
(M^{\prime }M)_{k}\thicksim (M^{\prime }M)_{t}. 
\]

If considering $\Delta t$ as a non-variable contradiction or a varied but
invariable contradiction, then as $(M^{\prime }M)_{t}=-\stackrel{\text{ }%
\cdot }{M}\Delta t$, the equation $aM^{\prime }M=-\stackrel{\text{ }\cdot }{M%
}\Delta t$ is evident due to the result of reflection 
\[
M^{\prime }M\thicksim -\stackrel{\text{ }\cdot }{M}\Delta t. 
\]
And therefore $a$ are generated to transfer worlds.

When taking $\Delta t$ into $a$ then we will obtain the equation of
causality, 
\begin{equation}
aM^{\prime }M=-\stackrel{\text{ }\cdot }{M}.  \tag{2.9}
\end{equation}

In the case the contradiction is characterized by itself, namely $M=M_{(M)}$%
, then 
\[
M=M_{0}e^{-a(t-t_{0})}, 
\]
where $M_{0}$ is the contradiction at the time $t=t_{0}$.

We next consider a case that if $\Delta t=[\top -\tau ]=0$ -- i.e. $\top
=\tau $, the time in the universe is identical in all spaces -- thereat the
time includes homogeneousness: it has an identical value everywhere at an
arbitrary point time of the universe, in the universe there is no deflection
of time at one place and other, at one space and other.

According to this significance, the time is a relatively independent world,
so whenever there is a time deflection then this deflection will must be
solved to reach a lowest possible deflection -- i.e. following the tendency $%
\Delta t=[\top -\tau ]\rightarrow 0$.

The time world stands of a background or a field, in which everything, every
system, every phenomenon and every state, etc. all reflect, and through
which to ``know'' differences in their motive process, to ``know''
differences between them and others and to find ways and tendencies for
solving.

Thus, if $\Delta t=[\top -\tau ]=0$, in the time world there is no
contradiction, then ${\it \Upsilon }$ will be equal to the unit $I$.
Therefrom, 
\[
f\hspace*{-0.1cm}!({\cal M}_{({\it \Upsilon })})=\stackrel{\text{ }\cdot }{%
{\cal M}}I. 
\]

And after soaking the universal equation into it, 
\[
f\hspace*{-0.1cm}!({\cal M}_{({\it \Upsilon })})={\cal M}^{\prime }{\cal M}=%
\stackrel{\text{ }\cdot }{{\cal M}}I, 
\]
with each functor element, 
\begin{equation}
A(M_{(t)})=e^{\Sigma aM_{(t)}\widehat{M}}M_{(t)}=M_{(t)}.  \tag{2.10}
\end{equation}

\section{CONCLUSION}

We have answered almost very difficult and mysterious questions of the
universe. Why do everything, every phenomenon, every system, every state,
every process and so on all perform and exhibit as what we know but not
perform and exhibit otherwise? Because, in very deep of things, phenomena
and processes there are different elements to annihilate mutually appearing
contradictions between them. Elements annihilate mutually leading gradually
to do not annihilate any more. This requirement makes them appear expedients
for solving of contradiction and facilities for transfer of contradiction in
the tendency of decreasing gradually contradictions between mutual
annihilation elements. And owing to variation of these contradictions as
well as of different elements in things, phenomena and processes, in the
universe there is an appearance of laws, in which everything, phenomenon,
process and so on all perform and exhibit just as that we see in the nature,
and anywhere in the universe. Then, why do only different elements play an
important role for performance and exhibition of everything, phenomenon and
process but do not any other elements? Because the difference is the first
axiom of the universe. If there were no existence of difference, then in the
universe everywhere all would be identical, and therefore none of anything
would exist. For that reason, only different elements are the first elements
and play an important role for exhibition and variation of things, phenomena
and processes in the universe. If so, then how is the universe in variation
and in motion?

Variation processes in the universe could not arise from condensed matter
and then create a big-bang generating particles, material bodies, etc. and
making the universe expand... Because the outset state of the universe must
be a state without immanent difference, or more exactly, with a least
quantity of immanent differences. If matter were condensed, then it would
comprise infinitely many different elements, therefore condensed state is
impossible to be the basic state. And if condensed matter is identical, then
in the universe there will not be the existence of a big-bang but matter
will must be varied in the expansion phase of the universal equation. In
reality, in the early universe infinitely many point spaces create
infinitely many ``big-bangs'' which do not explode out but burst into center
of each point.

Therefore, the universe varies and transforms in the process 
\[
f\hspace*{-0.1cm}!\rightarrow f\hspace*{-0.1cm}!(~)=0\rightarrow
f\hspace*{-0.1cm}!(K)\rightarrow \infty \rightarrow (1)\rightarrow
f\hspace*{-0.1cm}!. 
\]

In the debut, $f\hspace*{-0.1cm}!$ is equal gradually to zero everywhere
forming itself and $f\hspace*{-0.1cm}!(~)$, thereby the time world also is
gradually formed and contributes to crystallize the space world. The process
from $f\hspace*{-0.1cm}!$ to infinity is performed as in the universal
equation. After that, because every contradiction solved all reach a state
without difference -- i.e. everywhere, everything are all identical, and
finally lead to a homogeneous state. At that time the process from infinity
to $(1)$ is performed. The end is the process of $f\hspace*{-0.1cm}!$-zing
everywhere, closing a motion period.

Briefly, the whole universe is a unity entity in causal connections, all
phenomena seem to progress on intrinsic order following the universal
equation and the whole universe continually obeys a law -- the law of
causality. From observations in the microcosm as well as in the macrocosm
science always discovers deterministic ``disorders'' of the universe which
are controlled by a some ``intelligence'' -- due to contradiction.

\section*{Appendices}

\subsection*{A. Functor}

We identify that a function $f(K)$ expanded into a Taylor series has the
following form (with $\left. \frac{d}{dK}\right| _{k}\equiv \widehat{f}$) 
\[
f(K)=e^{(K-k)\widehat{f}}f(k), 
\]
or 
\[
f(K)=e^{\int_{k}^{K}\widehat{f}(\kappa )d\kappa }f(k). 
\]
This function is called functor.

It is clear that the functor may be sought from the equation of causality 
\[
\widehat{f}f=\frac{\partial f}{\partial K}. 
\]

For the multiplet series the functor $f(K_{1},K_{2},...)$ has the form as 
\[
f(K_{1},K_{2},...)=e^{\Sigma _{i}(K_{i}-k_{i})\widehat{f}%
_{i}}f(k_{1},k_{2},...), 
\]
or as a functional 
\[
f(K_{(z)})=e^{\int dz\ \delta K_{(z)}\widehat{f}(k,z)}f(k_{(z)}). 
\]

This function also may be sought from the equation of causality 
\[
\widehat{f}f(K)=\frac{\delta f(K)}{\delta K_{(z)}}. 
\]

Briefly, an arbitrary function can expand into the Taylor convergent series
then also can be written as a functor and it satisfies the equation of
causality. In reality, every function can expand into the Taylor series if
we recognize the existence of all zero-valued, finite-valued and
infinite-valued derivatives. At that time, in the series there is no
remainder and the series has infinite order number. Thus, every phenomenon
can be described in the form of functor, and variation and motion (i.e.
every process) of phenomena then can be expressed in the equation of
causality.

\subsection*{B. Life-like Mathematics}

The life-like mathematics is a new mathematical phrasing, in which objects
and quantities, etc. always are in motion. The life-like mathematics
describes relationships of equation-of-causality plurality and ``living'' of
the universal equation, etc. Therewithal, the life-like mathematics
comprises not only pictures of the nature but also abstract quantities with
copious proper lifetimes reflecting mutually in a unified common lifetime.

{\bf 1}! \ \ In the space of actions ${\cal K}$ there exists a set of free
actions 
\[
{\cal K}=\{K_{1},K_{2},...,K_{n},...\}. 
\]

\begin{description}
\item  {\it a}. \ If actions have not any common thing and they are alike,
then they are independent mutually 
\[
K_{1},K_{2},...,K_{n},...\equiv K_{1},K_{2},...,K_{n},... 
\]

\item  {\it b}. \ If actions have one element or many common elements, alike
elements between them, then they connect mutually by annihilation operations
``$-$''. If those actions are different from each other, then they
annihilate mutually generating contradictions:
\end{description}

\begin{itemize}
\item[1.]  $K_{i}-K_{i}-\cdots -K_{i}\equiv K_{i}.$

\item[2.]  If actions annihilate directly, then 
\[
K_{1}-K_{2}-\cdots -K_{n}=M, 
\]
with $n$ being less than number of actions in the space ${\cal K}$.
\end{itemize}

{\bf 2}! \ \ \ In the operator space ${\cal M}^{\prime }$ there exists a set
of free action operators: 
\[
{\cal M}^{\prime }=\{M_{1}^{\prime },M_{2}^{\prime },...,M_{n}^{\prime
},...\} 
\]
so that:

\begin{description}
\item  {\it a}. \ In the case where contradiction is characterized by itself 
$M=M_{(M)}$, then 
\[
M_{(M)}^{\prime }\equiv M^{(0)}=1. 
\]

\item  {\it b}. \ As operators act simultaneously onto contradiction: 
\begin{eqnarray*}
M_{i}^{\prime }+M_{i}^{\prime }+\cdots +M_{i}^{\prime } &\equiv
&M_{i}^{\prime }, \\
M_{1}^{\prime }+M_{2}^{\prime }+\cdots +M_{n}^{\prime } &\equiv &M^{\prime },
\end{eqnarray*}
with $n$ being less than number of operators in the space ${\cal M}^{\prime
} $. 
\[
\text{If \ \ \ }M_{2}^{\prime }=M_{-1}^{\prime }\text{, \ \ \ \ \ then \ \ \ 
}M_{1}^{\prime }+M_{2}^{\prime }=M_{1}^{\prime }+M_{-1}^{\prime }=(0)\text{,}
\]
(because they eliminate mutually.)

\item  ${\it c}$. \ As operators act onto contradiction in order: 
\[
M_{i}^{\prime }M_{i}^{\prime }...\equiv M_{i}^{\prime \prime \cdots }, 
\]
\[
M_{2}^{\prime }M_{1}^{\prime }...\equiv M_{2}^{\prime }M_{1}^{\prime }..., 
\]
\[
M_{2}^{\prime }M_{1}^{\prime }...=M_{-1}^{\prime }M_{1}^{\prime }=(0)\text{
\ \ \ \ \ \ \ if \ \ }M_{2}^{\prime }=M_{-1}^{\prime }, 
\]
(because $M_{1}^{\prime }$ and $M_{-1}^{\prime }$ eliminate mutually.)
\end{description}

{\bf 3}! \ \ \ Selection Rule.

\begin{description}
\item  {\it a}. \ As the two spaces of actions ${\cal K}_{1}$ and ${\cal K}%
_{2}$ sink into each other: each action in one space of actions searches and
selects actions among appropriate actions in other space in order to unite
mutually when between them there are common elements (if not, they have
mutually independent tendency) and then forming contradictions in a new
space ${\cal M}$: 
\[
|{\cal K}_{1}-{\cal K}_{2}|={\cal M}. 
\]
Writing so means that it has eliminated actions which do not contribute to
create contradictions.

\item  {\it b}. \ As the operator space ${\cal M}^{\prime }$ sink into the
space ${\cal M}$: each some contradiction $M_{i}$ in the space ${\cal M}$
searches and selects necessary operators (e.g. $M_{i}^{\prime }$) among
appropriate operators in the operator space ${\cal M}^{\prime }$ so that
those contradictions are solved with respect to transfer facilities (degrees
of freedom) which chosen operators have contained (if operators are not
necessary, contradictions will not choose them). These make the space ${\cal %
M}$ have contradiction parts to be solved, that solving obeys the equation
of causality and contradictions decrease gradually to reach a lowest
contradiction state. 
\[
aM_{i}^{\prime }M_{i}=-\stackrel{\text{ }\cdot }{M}_{i}. 
\]
Finally, the space ${\cal M}$ reaches and concludes at a new space of
actions ${\cal K}_{3}$.

\item  {\it c}. \ As the two operator spaces ${\cal M}_{1}^{\prime }$ and $%
{\cal M}_{2}^{\prime }$ sink into each other: then operators act onto each
other obeying the rule 
\[
M_{2}^{\prime }M_{1}^{\prime }=M_{1}^{\prime }M_{2}^{\prime }, 
\]
where $M_{1}^{\prime }\in {\cal M}_{1}^{\prime }$ and $M_{2}^{\prime }\in 
{\cal M}_{2}^{\prime }$.

\item  {\it d}. \ As the two spaces $({\cal M}^{\prime }{\cal M})_{1}$ and $(%
{\cal M}^{\prime }{\cal M})_{2}$ sink into each other: then the two spaces $%
{\cal M}_{1}$ and ${\cal M}_{2}$ sink into each other, and simultaneously
the spaces ${\cal M}_{1}^{\prime }$ and ${\cal M}_{2}^{\prime }$ sink into
the spaces ${\cal M}_{1}$ and ${\cal M}_{2}$.
\end{description}

{\bf 4}! \ \ \ Action Group.

There exist actions interacting together under the annihilation opearation $%
\ominus $ and action operators under the successive action operation $\cdot $
and simultaneous action operation $\oplus $. They obey the following laws:

\begin{description}
\item  {\it a}. \ In the space of actions $\{K\}$: there exist actions $%
K^{+} $ and anti-actions $K^{-}$; 
\begin{eqnarray*}
\lbrack K^{\pm }\ominus K^{\mp }] &=&0,\qquad \text{(existence of
anti-actions)} \\
\lbrack 0\ominus K^{\pm }] &=&K^{\pm },\qquad \text{(existence of unite }0%
\text{)} \\
\lbrack K_{i}^{\pm }\ominus K_{j}^{\pm }] &=&K_{k}^{\pm }\in \{K\},\qquad 
\text{(algebra)} \\
\lbrack K_{i}^{\pm }\ominus K_{j}^{\pm }] &=&[K_{j}^{\pm }\ominus K_{i}^{\pm
}],\qquad \text{(permutation)}
\end{eqnarray*}
\[
\lbrack \lbrack K_{i}^{\pm }\ominus K_{j}^{\pm }]\ominus K_{k}^{\pm
}]=[K_{i}^{\pm }\ominus \lbrack K_{j}^{\pm }\ominus K_{k}^{\pm
}]]=[[K_{k}^{\pm }\ominus K_{i}^{\pm }]\ominus K_{j}^{\pm }]. 
\]

\item  ${\it b}$. \ In the space of action operators $\{\widehat{M}\}$:
there exist action operators $M^{+\prime }$ and anti-action operators $%
M^{-\prime }$; 
\begin{eqnarray*}
M^{+\prime }\cdot M^{-\prime } &=&M^{-\prime }\cdot M^{+\prime }=1,\qquad 
\text{(existence of anti-action operators)} \\
M^{+\prime }\oplus M^{-\prime } &=&M^{-\prime }\oplus M^{+\prime }=1, \\
M^{\prime }\cdot M_{(M)}^{\prime } &=&M_{(M)}^{\prime }\cdot M^{\prime
}=M^{\prime }\oplus M_{(M)}^{\prime }=M_{(M)}^{\prime }\oplus M^{\prime
}=M^{\prime },\qquad
\end{eqnarray*}
(existence of unite $M_{(M)}^{\prime }=M^{(0)}=1$) 
\begin{eqnarray*}
M^{\prime }\oplus M^{\prime } &=&M^{\prime };\qquad M_{i}^{\prime }\oplus
M_{j}^{\prime }=M_{j}^{\prime }\oplus M_{i}^{\prime }=M_{k}^{\prime }\in \{%
\widehat{M}\}; \\
M^{\prime }\cdot M^{\prime } &=&M^{\prime \prime };\qquad M_{i}^{\prime
}M_{j}^{\prime }=\alpha _{_{ji}}M_{j}^{\prime }M_{i}^{\prime },\quad \alpha
_{ij}=\alpha _{_{ji}}^{-1}.
\end{eqnarray*}
\end{description}

Briefly, there are many more actions then many more such kinds of
contradictions and derivatives of contradictions and also the same for the
equation of causality.

Depending on concrete problem, the annihilation operation may be a vectorial
product, or may be a derivative limit, etc. and the degrees of freedom may
be a rotation angle, may be coordinates, or also may be the time, etc. and
similar to the world transfer coefficient may be a definite value depending
upon reflecting from one world to other in that concrete problem.

\section*{Acknowledgments}

We would like to thank Dr. D. M. Chi for useful discussions and valuable
comments.

\end{document}